\newif\ifcomment
\newif\ifdraft
\newif\ifTrigger
\newif\ifcolor
\newif\iflatexdiff
\def\dvers{v1.8}
\def\dtitle{Production of $\omega$ mesons in pp collisions at $\mathbf{\sqrt{s}=7\,\text{TeV}}$}
\def\stitle{Production of $\omega$ mesons in pp collisions at $\sqrt{s}=\SI{7}{TeV}$} 
\documentclass[ALICE,manyauthors]{cernphprep}
\usepackage[comma,square,numbers,sort&compress]{natbib}
\usepackage{hyperref}
\usepackage{graphicx}  % needed for figures
\usepackage{dcolumn}   % needed for some tables
\usepackage{bm}        % for math
\usepackage{amssymb}   % for math
\usepackage{amsfonts}
\usepackage{graphics}
\usepackage{grffile}   % to handle dots in graphics file names
\usepackage{epsfig}
\usepackage[loose]{units}
\usepackage[usenames,dvipsnames]{xcolor}
\usepackage[normalem]{ulem} % for strikethroughs (\sout{})
\usepackage{xcolor}
\usepackage[utf8]{inputenc}
\usepackage[T1]{fontenc}
\usepackage{subfigure}
\usepackage{mathrsfs}
\usepackage{rotating}
\usepackage{multirow}
\usepackage{booktabs}
\usepackage{lineno}
\usepackage[separate-uncertainty=true]{siunitx} % handle units nicely
\usepackage{physics} % makes life easier
\iflatexdiff
\RequirePackage{color}\definecolor{RED}{rgb}{1,0,0}\definecolor{BLUE}{rgb}{0,0,1}

\fi

\newcommand{\Pythia}       {\mbox{PYTHIA}}

\newcommand{\degree}       {\ensuremath^{\rm o}}

\newcommand{\pt}           {\ensuremath{p_{\mathrm{T}}}}
\newcommand{\pT}           {\pt}
\newcommand{\mT}           {\ensuremath{m_{\mathrm{T}}}}

\newcommand{\mt}           {\ensuremath{m_{\mathrm{T}}}}

\newcommand{\geant}        {{GEANT}}

\newcommand{\Fig}[1]       {Fig.~\ref{#1}}
\newcommand{\Figure}[1]    {Figure~\ref{#1}}
\newcommand{\Sect}[1]      {Sec.~\ref{#1}}
\newcommand{\Section}[1]   {Section~\ref{#1}}

\newcommand{\Eq}[1]        {Eq.~\ref{#1}}

\newcommand{\Tab}[1]       {Tab.~\ref{#1}}
\newcommand{\Table}[1]     {Table~\ref{#1}}
\newcommand{\Refs}[1]      {Refs.~\cite{#1}}

\newcommand{\com}[1]       {}
\iflatexdiff

\renewcommand{\xout}[1]    {\textcolor{red}{\sout{#1}}}
 
\newcommand{\old}[1]       {{\textcolor{red}{\sout{#1}}}}

\else

\renewcommand{\xout}[1]    {}
\newcommand{\old}[1]       {\relax}

\fi

\graphicspath{{./img/}}
\ifdraft
\usepackage{lineno}
\linenumbers
\modulolinenumbers[1]
\usepackage{fancyhdr}
\fi
\begin{document}
%==========================================================%
\begin{titlepage}
\PHyear{2020} %	CERN-EP-2020-122
\PHnumber{122}    % required, obtained from PH
\PHdate{16 June}    % required
\title{\dtitle}
\ShortTitle{\stitle}
\Collaboration{ALICE Collaboration%
         \thanks{See Appendix~\ref{app:collab} for the list of collaboration members}}
\ShortAuthor{ALICE Collaboration} % appears on left page headers, do not change
\begin{center}
\ifdraft
\today\\ \color{red}DRAFT \dvers\ \hspace{0.3cm} \$Revision: 6080 $\color{white}:$\$\color{black}\vspace{0.3cm}
\else
\fi
\end{center}
%==========================================================%
\begin{abstract}
The invariant differential cross section of inclusive $\omega(782)$ meson production at midrapidity~($|y|<0.5$) in pp collisions at $\sqrt{s}=\SI{7}{TeV}$ was measured with the ALICE detector at the LHC over a transverse momentum range of $2 < \pt < \SI{17}{GeV}/c$.
The $\omega$ meson was reconstructed via its $\omega\rightarrow\pi^+\pi^-\pi^0$ decay channel.
The measured $\omega$ production cross section is compared to various calculations: \Pythia~8.2 ~Monash 2013 describes the data, while \Pythia~8.2~Tune~4C overestimates the data by about 50\%.
A recent NLO calculation, which includes a model describing the fragmentation of the whole vector-meson nonet, describes the data within uncertainties below $\SI{6}{GeV}/c$, while it overestimates the data by up to 50\% for higher $\pT$.  
The $\omega/\pi^0$ ratio is in agreement with previous measurements at lower collision energies and the PYTHIA calculations.
In addition, the measurement is compatible with transverse mass scaling within the measured \pT~range and the ratio is constant with $C^{\omega/\pi^{0}}= 0.67  \pm 0.03 \text{~(stat)~} \pm 0.04 \text{~(sys)~}$ above a transverse momentum of $\SI{2.5}{GeV}/c$.
\end{abstract}
\end{titlepage}
\newpage
\setcounter{page}{2}
%==========================================================%
%==========================MAIN============================%
%==========================================================%
%%%%%%%%%%%%%%%%%%%%%%%%%%%%%%%%%%%%%%%%%%%%%%%%%%
\section{Introduction}
\label{sec:intro}
%%%%%%%%%%%%%%%%%%%%%%%%%%%%%%%%%%%%%%%%%%%%%%%%%%
Measurements of hadron production cross sections in proton-proton~(pp) collisions at high energies are important to test our understanding of strong interaction and its underlying theory of Quantum Chromodynamics~(QCD)~\cite{Gross:1973ju}. 
Its perturbative treatment~(pQCD) becomes feasible for predictions of particle production in hard scattering processes that have a sufficiently high momentum transfer $Q^2$.
This is possible by factorizing~\cite{Collins:1989gx} the scattering process into three contributions:
a QCD matrix element describing the scattering of partons, a parton distribution function~(PDF)~\cite{Collins:1981uw} describing the probability to find a scattering parton within each colliding hadron, and a fragmentation function~(FF)~\cite{Field:1977fa} that relates the final-state parton momentum to the momentum of an observed hadron.
While the QCD matrix element can be calculated in pQCD for sufficiently hard scales, the FFs and PDFs are obtained by global fits of experimental data at various collision energies~\cite{Gao:2017yyd}.
However, most particles are produced in soft scattering processes that involve small momentum transfers and therefore can not be calculated within pQCD.
In this regime, calculations rely on phenomenological models that also require experimental verification. 

Comparison of measured particle spectra with calculations is essential to test their underlying assumptions and provide constraints for the FFs and the PDFs.
For example, recent measurements of $\pi^0$ and $\eta$ mesons~\cite{Abelev:2012cn,Abelev:2014ypa,Acharya:2017tlv} at several LHC collision energies constrained gluon fragmentation~\cite{deFlorian:2014xna} in a regime not accessible by measurements at lower collision energies.
Like the $\pi^0$ and $\eta$ mesons, the $\omega$ meson is comprised mainly of light valence quarks and hence has similar flavor content.
However, it has spin 1 and is heavier than the $\pi^0$ and $\eta$ with a mass of $\SI{782}{MeV}/c^2$~\cite{Tanabashi:2018oca}.
These differences make the $\omega$ meson an interesting complementary probe to improve our understanding of hadron production in high-energy collisions.
Even though there have been several theoretical efforts to describe the fragmentation into pseudoscalar mesons and baryons such as $\pi$, K, $\eta$ and protons \cite{Albino:2008gy,Bertone:2017tyb}, only a few theoretical models exist to describe the fragmentation into vector mesons, due to a lack of experimental data.
Nonetheless, recent efforts~\cite{Saveetha:2013jda,Saveetha:2017xmc} have been made to describe the fragmentation  into the entire vector meson nonet using a model with broken SU(3) symmetry by analysing RHIC (pp) and LEP ($e^+e^-$) data.

This article presents the invariant differential cross section of inclusive $\omega$ meson production at mid-rapidity~($|y|<0.5$) in pp collisions at $\sqrt{s}=\SI{7}{TeV}$.
The cross section of $\omega$ production in hadronic interactions has been measured at collision energies of $\sqrt{s}=\SI{62}{GeV}$~\cite{ISRomega} and $\sqrt{s}=\SI{200}{GeV}$~\cite{Adare:2010fe,RHICOmega1,RHICOmega2} at ISR and RHIC respectively.
At LHC energies, $\omega$ production has only been measured by ALICE at forward rapidities ($2.5<y<4.0$) in pp collision at \SI{7}{TeV}~\cite{ALICE:2011ad} in a transverse momentum~($\pt$) range of $1<$~\pT~$<\SI{5}{GeV}/c$.
The results reported here provide the first measurement of $\omega$ production at mid-rapidity at LHC energies, and in a wide \pt~range of $2<\pt<\SI{17}{GeV}/c$, which tests existing calculations in this regime and provides input for future theoretical studies of vector meson fragmentation functions.
In addition, the $\omega/\pi^0$ production ratio as a function of \pt~is compared to results of measurements at lower collision energies.
This ratio also tests the validity of transverse mass~($\mt$) scaling~\cite{Altenkamper:2017qot} for $\omega$ mesons at LHC energies, which is typically applied to estimate hadronic backgrounds in direct photon or di-electron measurements in situations where no measured hadron spectra are available. 
The empirical scaling rule, which was established in measurements of identified particle spectra at lower collision energies at ISR and RHIC~\cite{Khandai:2011cf}, states that the $\pt$-differential yields of most particles can be described as $E\dv*[3]{\sigma}{p}=C^h f(\mt)$, where $f(\mt)$ is a universal function for all hadron species and $C^h$ is a constant normalisation factor.

The article is structured as follows:
\Section{sec:alicedetector} briefly describes the ALICE sub-detectors, with a focus on those relevant for the measurement.
Details on the event selection and signal extraction are given in Secs.~\ref{sec:eventtrack}  to \ref{sec:meson}.
Sources of systematic uncertainties are discussed in \Sect{sec:systematics}.
The data and comparisons to model predictions are presented in \Sect{sec:results}.
Finally, conclusions are provided in \Sect{sec:conclusion}.
%%%%%%%%%%%%%%%%%%%%%%%%%%%%%%%%%%%%%%%%%%%%%%%%%%
\section{ALICE detector}
\label{sec:alicedetector}
%%%%%%%%%%%%%%%%%%%%%%%%%%%%%%%%%%%%%%%%%%%%%%%%%%
The $\omega$ meson was reconstructed via its decay to $\pi^+\pi^-\pi^0$, where in turn the $\pi^0$ decays to two photons.
This strategy required the reconstruction of charged tracks in the ALICE central tracking system, composed of the Inner Tracking System~(ITS)~\cite{Dellacasa1999kf} and the Time Projection Chamber~(TPC)~\cite{Alme:2010ke}, and the reconstruction of photons using the ElectroMagnetic Calorimeter~(EMCal)~\cite{Cortese:2008zza,Abeysekara:2010ze} and the Photon Spectrometer~(PHOS)~\cite{Dellacasa:1999kd}.
In addition, photons were reconstructed using the Photon Conversion Method (PCM) \cite{Abelev:2014ffa}, which exploits the capability of the central tracking system to reconstruct photons from electron-positron track pairs.
A detailed description of the ALICE detector system and its performance can be found in \Refs{Aamodt:2008zz} and~\cite{Abelev:2014ffa}, respectively.
Below, a brief overview of the previously mentioned detectors and the V0 detector~\cite{Abbas:2013taa}, used for the minimum bias trigger, is given.

The ITS is positioned closest to the nominal interaction point and consists of two layers of Silicon Pixel Detectors~(SPD), two layers of Silicon Drift Detectors~(SDD) and two outermost layers of Silicon Strip Detectors~(SSD).
The layers are positioned between \SI{3.9}{cm} and \SI{43.0}{cm} radial distance from the beamline, where the two SPD layers cover a pseudorapidity range of $|\eta|<2$ and $|\eta|<1.4$, respectively.
The SDD and SSD have a pseudorapidity coverage of $|\eta|<0.9$ and $|\eta|<1.0$, respectively.
The ITS is used for the tracking of charged particles and the reconstruction of the primary vertex.

The TPC is a large~($\SI{90}{m^3}$) cylindrical drift detector, which allows for the measurement of charged particles and their identification via specific energy loss~($\dv*{E}{x}$) measurements.
The TPC covers a pseudorapidity range of $|\eta|<0.9$ over the full azimuth and enables the measurement of up to 159 space points per track.
A large solenoidal magnet surrounding the central barrel detectors provides a magnetic field of $B=\SI{0.5}{\tesla}$, which allows one to reconstruct tracks down to $\pt\approx\SI{100}{MeV}/c$. For the reconstruction of charged particles in the ITS and TPC, a transverse momentum resolution of about $1\%$ at $\SI{1}{GeV}/c$ is achieved, which decreases to about $3\%$ at $\SI{10}{GeV}/c$~\cite{Alme:2010ke}.

The EMCal is a Pb-scintillator sampling calorimeter, which covered an azimuthal range of $\Delta\varphi=\SI{40}{\degree}$ and $|\eta|<0.67$ in pseudorapidity during 2010 data taking. In that period, it was comprised of 4 super modules, each consisting of 288 modules.
The module consists of four towers with a size of $\approx 6\times\SI{6}{cm^2}$, corresponding to approximately twice the Moli\`ere radius. Each tower is made up of 140 alternating lead and scintillator layers, where the latter are connected to Avalanche Photo Diodes (APDs) that measure the scintillation light of the electromagnetic showers produced by particles traversing the lead absorber. 
The energy resolution is given by $\sigma_E/E=4.8\%/E\oplus 11.3\%/\sqrt{E}\oplus1.7\%$ with energy $E$ in units of GeV~\cite{Abeysekara:2010ze}. 

The PHOS is an electromagnetic calorimeter with high granularity based on lead-tungstate~(Pb$\text{WO}_4$) scintillation crystals.
At the time these data were collected, it had an acceptance of $\Delta\varphi=\SI{60}{\degree}$ and $|\eta|<0.12$, divided into three modules, each consisting of 3584 crystals that are connected to APDs. % and a low-noise preamplifier.
A high granularity is achieved by small crystal size of $\approx 2.2\times \SI{2.2}{cm^2}$, where the lateral dimensions of the cells are only slightly larger than the Pb$\text{WO}_4$ Moli\`ere radius of \SI{2}{cm}.
The high light yield of the Pb$\text{WO}_4$ crystals operated at \SI{-25}{\degreeCelsius} results in an energy resolution of $\sigma_E/E=1.3\%/E\oplus3.6\%/\sqrt{E}\oplus1.1\%$ with energy $E$ in units of GeV~\cite{Acharya:2019rum}.

The V0 detector provides the minimum bias triggers and is employed to reduce background events, such as beam-gas interactions and out-of-bunch pileup.
It consists of two scintillator arrays located in the forward and backward rapidity regions of the ALICE apparatus, covering a pseudorapidity of $2.8<\eta<5.1$ and $-3.7<\eta<-1.7$, respectively.
%%%%%%%%%%%%%%%%%%%%%%%%%%%%%%%%%%%%%%%%%%%%%%%%
\section{Event and track selection}
\label{sec:eventtrack}
%%%%%%%%%%%%%%%%%%%%%%%%%%%%%%%%%%%%%%%%%%%%%%%%
The pp collision data used for the $\omega$ meson measurement were recorded by the ALICE experiment in 2010 \ifTrigger and 2011\fi at a centre-of-mass energy of $\sqrt{s}=\SI{7}{TeV}$.
In 2010, a minimum bias trigger $\text{MB}_{\text{OR}}$, which required a signal either in the SPD or in one of the V0 scintillator arrays, was used.
The total inelastic pp collision cross section was determined on the basis of the van der Meer scan and was found to be $\sigma_{\text{inel}}=73.2^{+2.0}_{-4.6}\text{(model)}\pm 2.6\text{(lumi)}\,\si{\milli\barn}$~\cite{Abelev:2012sea}.
The corresponding cross section of the $\text{MB}_{\text{OR}}$ trigger was $\sigma_{\text{MB}_{\text{OR}}}=\SI{62.4\pm2.2}{\milli\barn}$. 
\ifTrigger
In 2011, the data were sampled by the EMCal Level-0~(L0) trigger, were an energy deposition above \SI{5.5}{GeV} in a $2\cross2$ tower window is required in addition to the minimum bias trigger condition $\text{MB}_{\text{AND}}$, requiring at least one hit in V0A and V0C.
\fi
Beam-induced background events, such as beam-gas interactions or out-of-bunch pileup, are rejected offline by using the timing information from the V0 detectors and the number of reconstructed hit points and track segments in the SPD, which are expected to be uncorrelated for background events.
The rejection of in-bunch pileup events, where multiple interactions occur per bunch crossing, was achieved by requiring that only a single primary vertex is reconstructed with the SPD per event.
Moreover, collision events with a reconstructed vertex more than \SI{10}{cm} away from the nominal interaction point along the beam axis were rejected.
The integrated luminosities $\mathcal{L}_{\text{int}} = N_{\text{evt}}/ \sigma_{\text{MB}_{\text{OR}}}$ were determined to be $\mathcal{L}_{\text{int}}^{\text{EMCal}}=\SI{6.4\pm 0.2}{\nano\barn^{-1}}$ and $\mathcal{L}_{\text{int}}^{\text{PHOS}}=\SI{6.0\pm 0.2}{\nano\barn^{-1}}$ for the measurement involving the EMCal and PHOS, respectively.
The integrated luminosity of the sample using only the PCM for photon reconstruction amounts to $\mathcal{L}_{\text{int}}^{\text{PCM}}=\SI{7.4\pm 0.3}{\nano\barn^{-1}}$.

Charged pion trajectories (tracks) with $|\eta|<0.9$ were reconstructed in the ITS and TPC, requiring at least 70 crossed cathode pad rows in the TPC and at least one hit in any of the layers of the ITS.
Furthermore, the $\chi^2$ of the track refit procedure per TPC space point was required to be below 4 and tracks with a momentum below $\SI{100}{MeV}/c$ were rejected. % due to their large curvature in the TPC.
The tracks were loosely constrained to the collision vertex by requiring a maximum distance of closest approach of a few centimeters to the collision vertex in beam direction and transverse plane. The resolution of the transverse distance to the primary vertex for ITS and TPC charged particle tracks is below \SI{150}{\micro m} for \pT$\gtrsim\SI{0.5}{GeV}/c$ ~\cite{Abelev:2014ffa}. 
Furthermore, charged pions can be identified using the specific energy loss $\dv*{E}{x}$ along the track in the TPC~\cite{Adam:2015qaa}. 

%%%%%%%%%%%%%%%%%%%%%%%%%%%%%%%%%%%%%%%%%%%%%%%%
\section{Photon measurement}
\label{sec:photon}
%%%%%%%%%%%%%%%%%%%%%%%%%%%%%%%%%%%%%%%%%%%%%%%%
To enhance the probability of the reconstruction of $\pi^0$ mesons, all methods to measure photons and $\pi^0$s at midrapidity with ALICE were exploited.
The EMCal and the PHOS allow for the measurement of photons via their electromagnetic shower deposits above \SI{\sim0.5}{GeV}, while the PCM enables the measurement of photons down to lower $\pT$ by exploiting the $e^+e^-$ pair creation by a photon within the inner detector material. 
Looser photon selection criteria as in previous publications, see e.g.~Ref.~\cite{Acharya:2017hyu}, were applied to increase the $\omega$ reconstruction efficiency.

The electromagnetic shower produced in the EMCal or PHOS by an incoming particle usually spreads over multiple adjacent towers, requiring the combination of the individual energy depositions to so-called clusters, which is achieved by clusterisation algorithms~\cite{Abelev:2014ffa}.
Each reconstructed cluster in the EMCal and PHOS was required to have a total energy of $E_{\text{clus}}>\SI{0.7}{GeV}$ and $\SI{0.3}{GeV}$ respectively to suppress contributions from minimum-ionising particles and noise. Additionally, in case of the EMCal, it ensures a good timing resolution. 
Detector noise in a single tower was removed by only selecting clusters with at least $2$ (EMCal) or $3$ (PHOS) towers for analysis.
In order to remove clusters from pileup events originating from subsequent bunch crossings, which occur in $\approx\SI{150}{ns}$ intervals, a cut on the timing of the leading tower for EMCal clusters of $-\SI{100}{ns} < t_{\text{cluster}} < \SI{100}{ns}$ with respect to the collision time was applied. 
Photon clusters were selected according to their cluster shape and, additionally, a track-matching procedure was applied to suppress clusters originating from charged particles reconstructed in the tracking system.
The EMCal cluster shape is parametrised by  the larger eigenvalue $\sigma_{\text{long}}^2$ of the dispersion matrix of the shower shape ellipse~\cite{Acharya:2017hyu,Awes:1992yp}.
A requirement of $0.1\leq \sigma_{\text{long}}^2 \leq 0.5$ was imposed, where the lower threshold removes contamination from non-physical background.
The upper threshold suppresses elongated clusters originating from low-\pt~electron and hadron tracks that hit the calorimeter surface not perpendicularly and merged clusters. The latter mostly originate from high-\pt~neutral pions that decay with a small opening angle, resulting in both decay photons to be reconstructed as a single cluster.

Photons traversing the detector material of ALICE convert to an electron-positron pair with a probability of about 8.5\% \cite{Abelev:2014ffa} within a radial distance of \SI{180}{cm} from the beam axis.
Such photons can be reconstructed using the PCM, which allows for the measurement of photons converting in the ITS and TPC within the fiducial acceptance of $|\eta|<0.9$.
First, secondary vertices~($\text{V}^0$s) were reconstructed by an algorithm~\cite{Alessandro:2006yt} exploiting the distinct topology of two tracks with opposite curvature that originate from a common point within the tracking detectors.
Good reconstruction quality of the tracks associated with a secondary vertex was assured by requiring $\pt>\SI{50}{MeV}/c$ and for the track to be comprised of at least 60\% of the findable TPC clusters.
Tracks originating from electrons were identified via their specific energy loss $\dv*{E}{x}$ in the TPC, which was required to be within $-3$ to $5$ $\sigma_e $~of the expected energy loss of electrons, where $\sigma_e$ is the standard deviation of the measured $\dv*{E}{x}$ distribution of electrons.
Contamination of charged pion tracks was suppressed by rejecting tracks whose $\dv*{E}{x}$ was within $\pm1\sigma_{\pi^{\pm}}$ of the expected energy loss for pions.
Several additional selection criteria were applied to identify $\text{V}^0$ candidates originating from photon conversions, exploiting the kinematics and topology of the conversion, as discussed in more detail in Ref.~\cite{Acharya:2017tlv}.
These include, e.g. selections to assure that the momentum vector of a conversion pair is pointing towards the primary vertex and a selection based on the minimal distance between the conversion point and the primary vertex, in order to remove contributions from Dalitz decays.
Furthermore, the quality of the obtained $\text{V}^0$ candidates was improved by constraining the reduced $\chi^2$ of the Kalman-filter hypothesis for the track pair.
Remaining contamination from $\mathrm{K}_{\text{S}}^0$, $\Lambda$ and $\bar{\Lambda}$ decays was reduced by a selection based on the decay kinematics in an Armenteros-Podolanski plot~\cite{podolanski1954analysis}, where photon conversions contribute as symmetric decays of a particle with vanishing rest mass.
Compared to previous PCM measurements~\cite{Acharya:2017tlv,Acharya:2017hyu}, a \pT~dependence of the selection criteria was introduced to further reduce the contamination from $\text{K}_{\text{S}}^0$ and $\Lambda$ decays.

\begin{figure}[t!]
	\centering
	\includegraphics[width=0.32\linewidth]{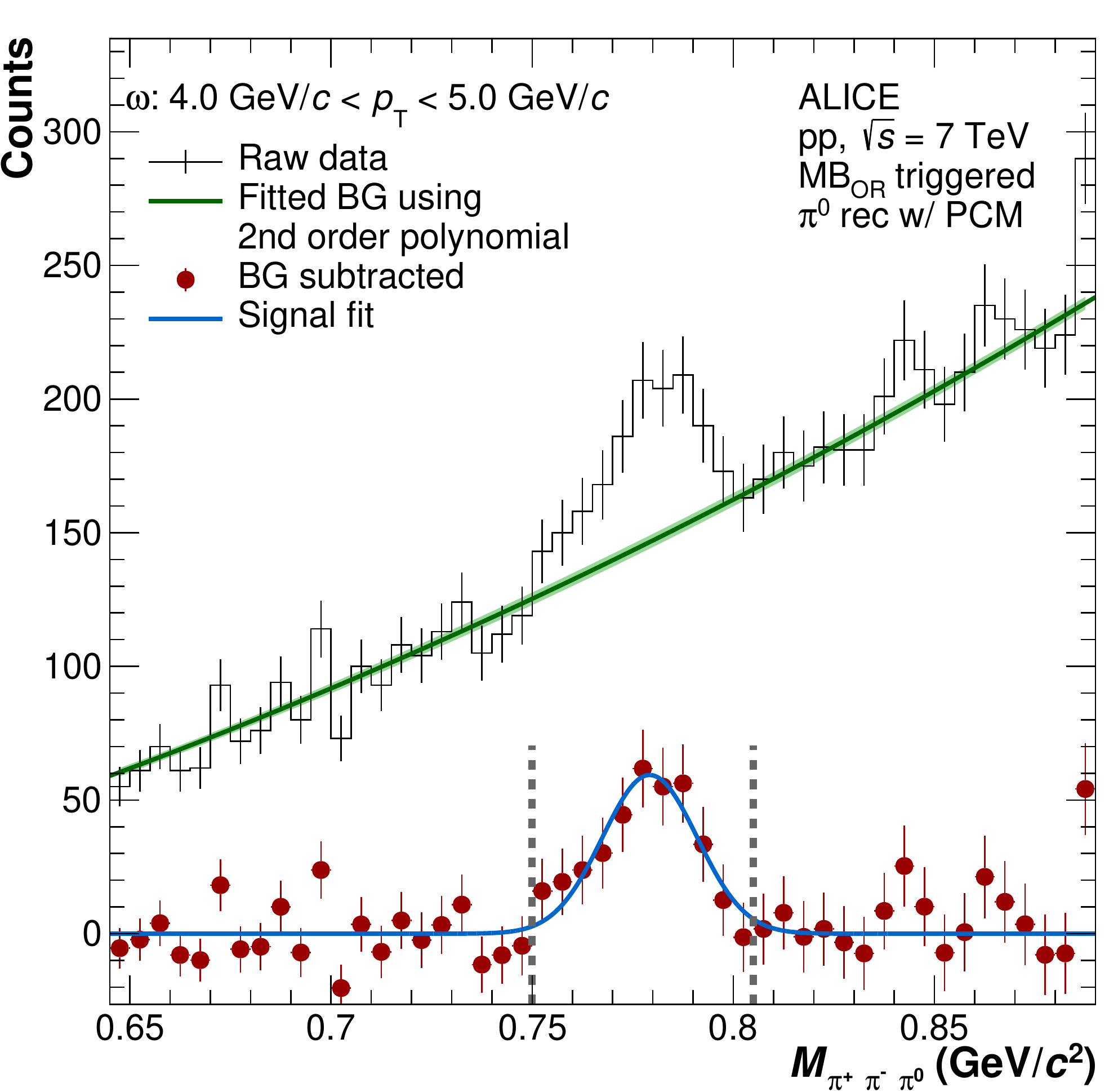}
	\includegraphics[width=0.32\linewidth]{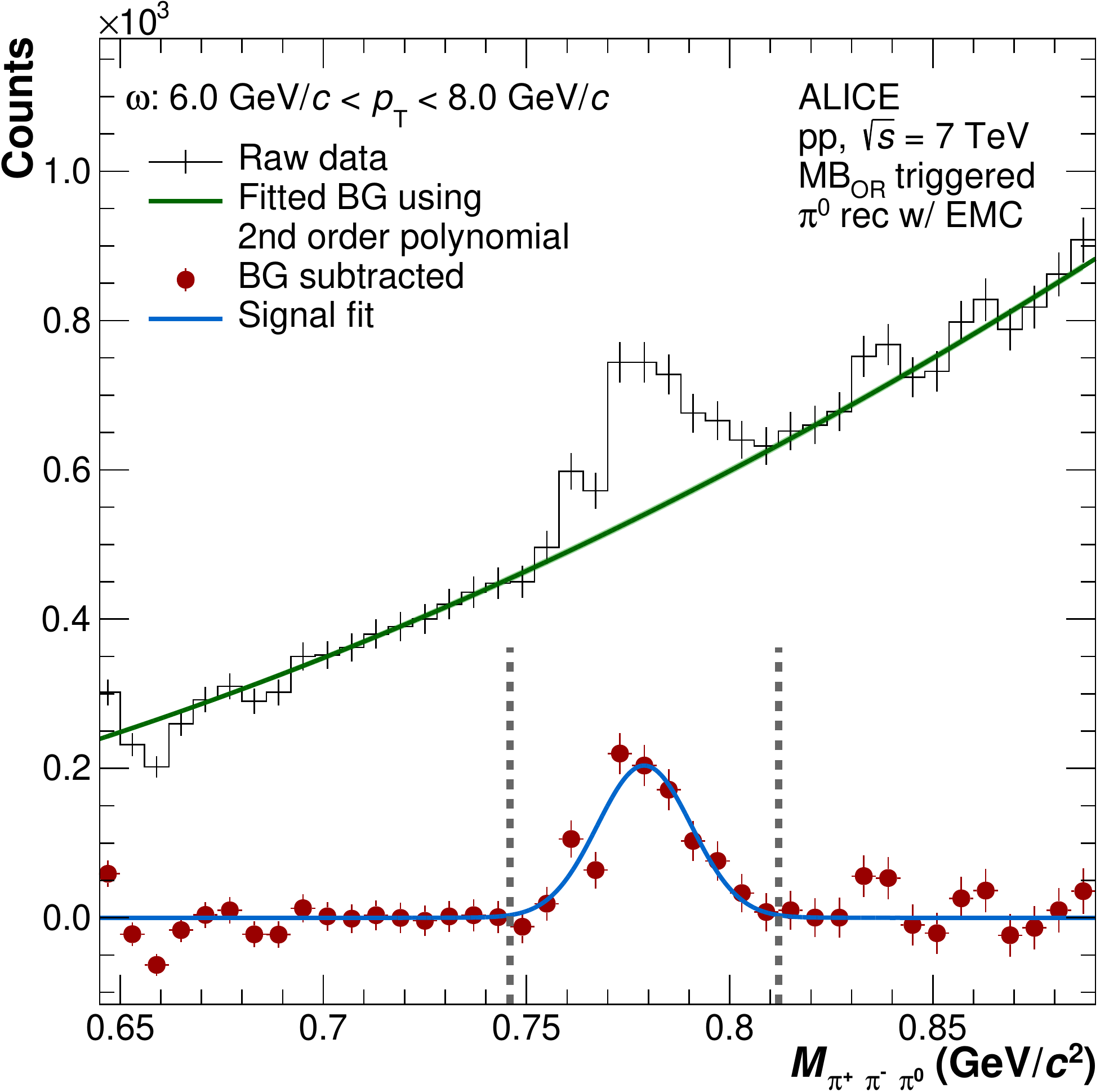}
	\includegraphics[width=0.32\linewidth]{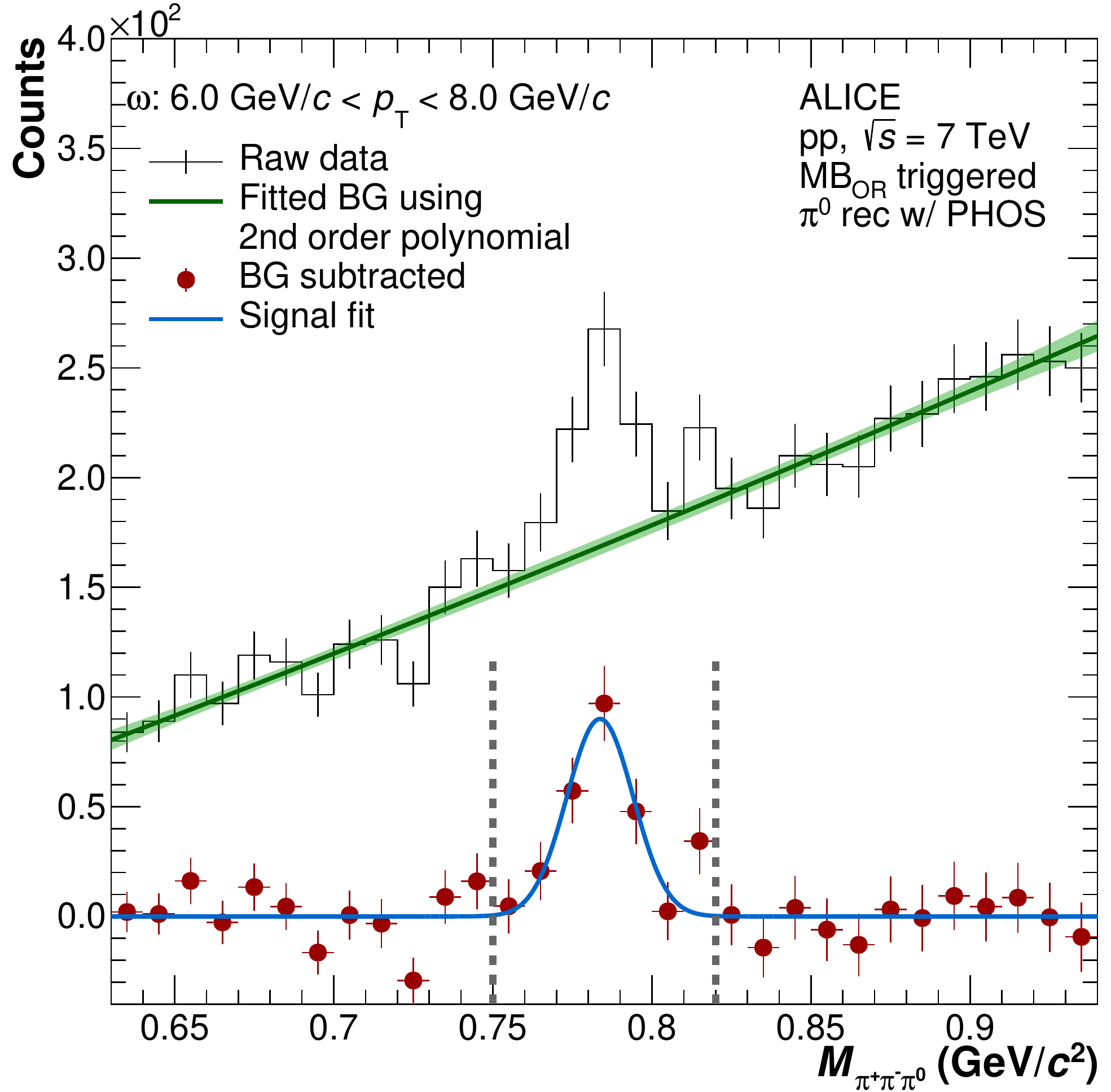}
	\caption{Invariant mass of $\pi^+\pi^-\pi^0$ candidates shown in the vicinity of the nominal mass of the $\omega$ meson for indicated \pT-ranges for $\pi^0$ reconstruction with PCM~(left), EMC~(middle) and PHOS~(right).
          The second order polynomial used for the background description is shown with a band denoting the statistical uncertainties of the fit.
          The points show the signal obtained after subtraction of the background fit.
          The signal is fitted with a Gaussian, where the vertical lines indicate the integration range used to obtain the raw yield by bin-by-bin counting, as outlined in \Sect{sec:meson}.}
	\label{omegaexamplebins}
\end{figure}

%%%%%%%%%%%%%%%%%%%%%%%%%%%%%%%%%%%%%%%%%%%%%%%%
\section{Meson reconstruction}
\label{sec:meson}
%%%%%%%%%%%%%%%%%%%%%%%%%%%%%%%%%%%%%%%%%%%%%%%%
In order to reconstruct the $\omega$ mesons via their $\pi^+\pi^-\pi^0$ decay channel, where the $\pi^0$ decays to two photons with a branching ratio of  $\approx 99\%$, a prior selection of $\pi^0$ candidates from pairs of photon candidates was applied.
For the photons that passed the selection criteria, the two-photon invariant mass ($M_{\gamma\gamma}$) of all possible photon pairs in a given event was calculated.
Four different methods were used for the $\pi^0$ candidate reconstruction, differing in how the photons entering the $M_{\gamma\gamma}$ calculation were selected.
These are referred to as PCM, PHOS and EMC, when \emph{both} photons used for the $\pi^0$ reconstruction were measured with the respective method.
In addition, a hybrid method~(PCM-EMC) was used, where one PCM photon was combined with a photon measured with the EMCal.
The resulting invariant mass distributions exhibit a peak of photon pairs originating from $\pi^0$ decays on top of combinatorial background.
The peak was parametrised in \pT~slices with a Gaussian to characterize the mean and width~($\sigma_{\pi^0}$) of the $\pi^0$ mass distribution. 
Photon pairs lying within about $\pm2\sigma_{\pi^0}$ of the expected $\pi^0$ mass were selected as neutral pion candidates for the $\omega$ meson reconstruction.
For the PHOS measurement~\cite{omegaphosPN}, $\pi^0$ candidates were furthermore required to have both photons in the same PHOS module and to have a minimum transverse momentum of $\SI{1.5}{GeV}/c$.
Finally, the nominal neutral pion mass was assigned to the mass of selected $\pi^0$ candidates in order to improve the $\omega$ mass resolution. 
This was achieved by subtracting the difference between the reconstructed $\pi^0$ mass and its nominal mass from the reconstructed $\omega$ mass.

Analogously to the $\pi^0$ reconstruction, the invariant mass of all $\pi^+\pi^-\pi^0$ combinations in a given event was determined by summing the four-momentum vectors of the candidate decay products passing the selection criteria. While charged pions were identified by requiring a $\dv*{E}{x}$ within $\pm3\sigma$ of their expected energy loss, no such selection was applied for the $\omega$ analysis with the $\pi^0$ reconstructed in PHOS to improve the $\omega$ reconstruction efficiency.

\Figure{omegaexamplebins} shows the invariant mass distribution in the vicinity of the $\omega$ nominal mass for indicated \pt~intervals for the $\pi^0$ reconstructed with PCM, EMC and PHOS, where a peak originating from $\omega$ meson decays is clearly visible above the combinatorial background.
The latter can be described using a second order polynomial for $\pT <\SI{10}{GeV}/c$. 
At higher momenta, a first order polynomial was used for the PHOS measurement.
The signal obtained after background subtraction was fitted with a Gaussian and the raw yield was obtained by adding counts within $\pm 2\sigma_{\omega}$~($\pm 3\sigma_{\omega}$ for PHOS) of the reconstructed $\omega$ mass, where $\sigma_{\omega}$ denotes the standard deviation of the Gaussian $\omega$ signal fit. 
The $\omega$ mass resolution was found to be about $\SI{15}{MeV}/c^2$ with a slight dependence on $\pT$  and reconstruction technique. 
This is achieved by the use of the previously mentioned nominal mass assignment for $\pi^0$ candidates, which improved the mass resolution by up to $30\%$.

\begin{figure}[t!]
	\centering
	\includegraphics[width=0.5\linewidth]{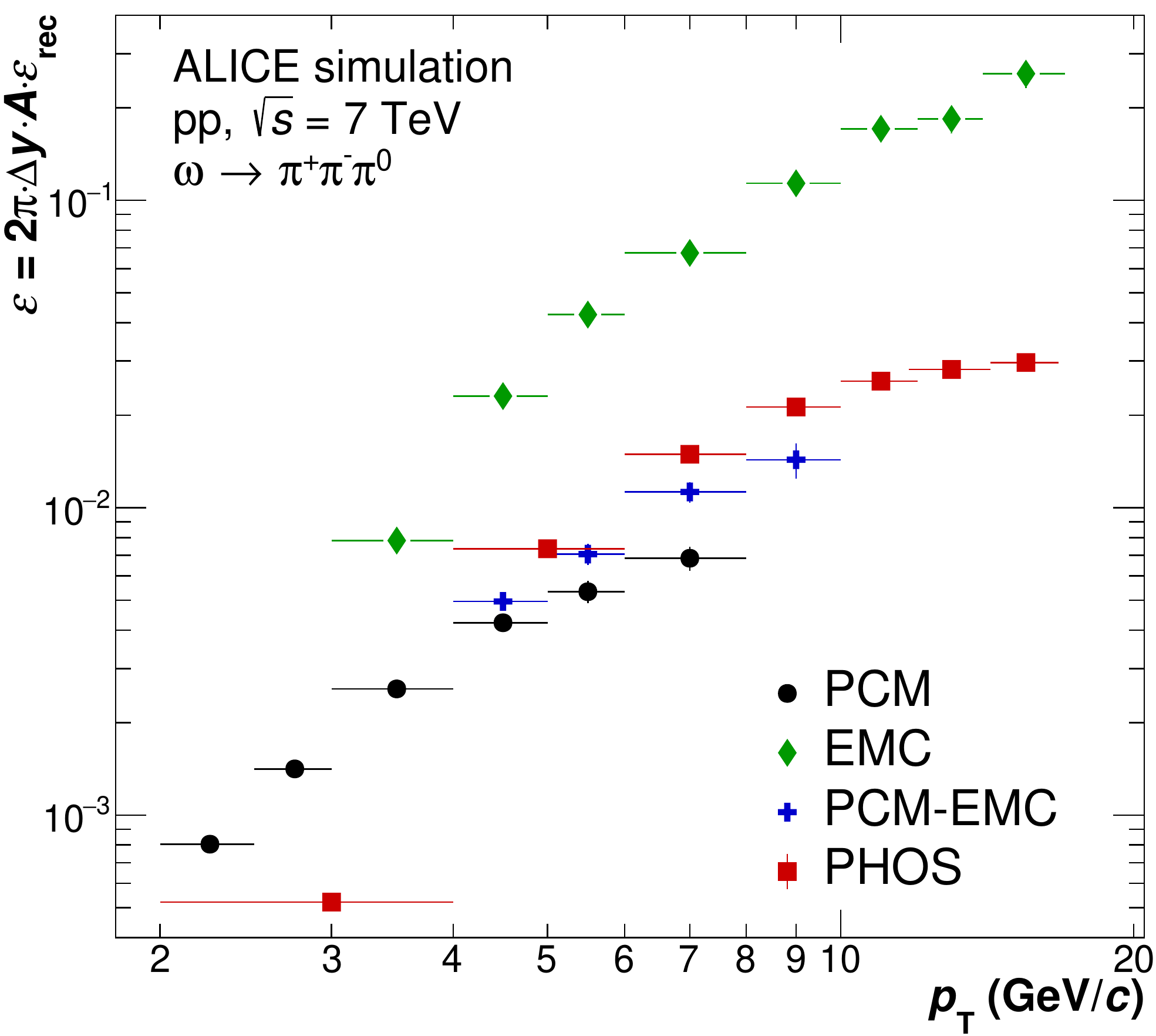}
	\caption{Correction factors applied to the raw $\omega$ yields for each indicated $\pi^0$ reconstruction method.
          The factors include the geometrical acceptance $A$ and the reconstruction efficiency $\epsilon_{\text{rec.}}$. In addition, a normalisation to unit rapidity and $2\pi$ azimuth angle is applied to allow for a direct comparison between the different methods.}
	\label{correctionfactor}
\end{figure}

The obtained raw yields for each reconstruction method were corrected for geometrical acceptance and reconstruction efficiency, which were evaluated using Monte Carlo simulations.
The event generator \Pythia 6.2~\cite{Sjostrand:2006za} was used to simulate the minimum bias pp collisions, where the implemented kinematics of the $\omega$ three-body decay are weighted assuming the experimentally observed phase space density distributions \cite{Stevenson:1962zz,Danburg:1971ui}.
All final state particles were propagated through the ALICE detector using \geant~3~\cite{geant3ref2}, taking into account the operating conditions of the detector at the time of data taking.
For each calorimeter, PHOS and EMCal, the relative difference in the energy scale and the non-linearity were tuned in the Monte Carlo to ensure agreement between the \pT-dependent reconstructed $\pi^0$ mass and width in data. 
This agreement propagates to the $\omega$ candidates, where mass and width in data and Monte Carlo are found to be consistent within the statistical uncertainties.
The full correction factors, $\varepsilon$, that were applied to the raw yields for the four different methods are shown in \Fig{correctionfactor}.
These factors include the geometrical acceptance evaluated for each method and the reconstruction efficiency, where the former is normalised to unit rapidity and $2\pi$ azimuth angle to allow for a direct comparison between the different methods.
The use of the four reconstruction techniques combines the strengths of the individual methods and maximizes the accessible \pT~reach. 
The reconstruction with PCM offers a low \pt-reach, however, the efficiency is limited due to the low conversion probability of about 8.5\%, while the reconstruction with the two calorimeters complements the measurement at high \pT.

%%%%%%%%%%%%%%%%%%%%%%%%%%%%%%%%%%%%%%%%%%%%%%%% 
\section{Systematic uncertainties}
\label{sec:systematics}
%%%%%%%%%%%%%%%%%%%%%%%%%%%%%%%%%%%%%%%%%%%%%%%%
The systematic and statistical uncertainties on the measured $\omega$ yield for the four individual reconstruction techniques in exemplary \pt\ intervals are summarised in \Tab{tab:table_system_Pi0}.
The uncertainties are given as relative uncertainties of the measured values in percent.

\begin{table}[thb!]
	\begin{center}
		              \caption{Overview of the relative uncertainties given in percent in exemplary \pT-intervals for the four individual reconstruction techniques of the $\omega$ meson.
			The given categories summarise systematic uncertainties arising from each analysis step.
			For each method the statistical and total uncertainties are reported in addition, as well as the uncertainties of the combined measurement. The uncertainty from the $\sigma_{\text{MB}_{\text{OR}}}$ determination of 3.5\% is independent from the individual measurements and indicated separately in Fig.~\ref{crosssection}.
		}
		\label{tab:table_system_Pi0}
		\scalebox{0.88}{
			\hspace*{-0.5cm}
			\begin{tabular}{l||c|c|c||c|c|c|c||c|c}
				\hline
				\pT~interval & \multicolumn{3}{c||}{$4-5$ GeV/$c$} & \multicolumn{4}{c||}{$6-8$ GeV/$c$} & \multicolumn{2}{c}{$12-14$ GeV/$c$}  \\ \hline
				\multirow{2}{*}{Method} & \multirow{2}{*}{PCM} & PCM- & \multirow{2}{*}{EMC}  & \multirow{2}{*}{PCM} & PCM- & \multirow{2}{*}{EMC} & \multirow{2}{*}{PHOS} &  \multirow{2}{*}{EMC} &\multirow{2}{*}{PHOS} \\ 
				&  & EMC  & & & EMC & & & & \\ \hline\hline
				Signal extraction      & $12.3$ & $12.6$ & $12.2$ &  $13.5$ & $13.5$ & $12.3$ & $6.0$ & $18.9$ &$11.0$ \\
				Material         &  $9.0$ & $4.7$ & $3.0$ &   $9.0$ & $4.7$ & $3.0$ & $3.5$  & $3.0$ & $3.5$ \\
				Charged pion rec.        & $6.8$ & $6.8$ &  $6.8$  &  $6.8$ & $6.8$ &  $6.8$ & $6.0$ & $6.8$ & $6.0$ \\
				Conv photon rec.      & $4.1$ & $4.1$ &     -  & $4.1$ & $4.1$ &     - &     - & - & -\\
				Calo photon rec.      & - & $5.0$ &  $6.9$    & - & $5.0$ &   $6.9$ &  $5.2$ & $6.9$& $9.3$\\
				Neutral pion rec.         & $6.0$ & $6.0$ &  $6.0$  & $6.0$ & $6.0$ &  $6.0$ &   $4.0$ &$6.0$&$ 4.0$ \\
				Pileup            &    $0.5$ & $0.5$ & $0.5$ &     $0.5$ & $0.5$ & $0.5$ & 0.5 & $0.5$&$0.5$\\
			 \hline
				Total syst. uncertainty& $18.1$& $17.7$ & $17.0$  & $19.1$& $18.3$ & $17.1$ & $11.0$ &$22.4$ &$16.0$ \\ 
				\hline\hline
				Statistical uncertainty& $14.5$ & $14.7$ & $9.8$& $18.9$ & $22.0$ & $9.2$ & $13.0$ & $21.7$ & $32.0$ \\
				\hline\hline
				Combined stat. unc.    & \multicolumn{3}{c||}{7.4} & \multicolumn{4}{c||}{7.2} & \multicolumn{2}{c}{18.0} \\
				Combined syst. unc.    & \multicolumn{3}{c||}{13.7} & \multicolumn{4}{c||}{10.3} & \multicolumn{2}{c}{16.6} \\ 
				\hline
		\end{tabular}}
	\end{center}
\end{table}

The signal extraction dominates the systematic uncertainties of the measurement and includes uncertainties due to the yield extraction.
For the PCM, PCM-EMC and EMC techniques the yield extraction uncertainty was estimated by varying simultaneously the bin-counting window used to obtain the raw yield in data and Monte Carlo and the fit range used for the polynomial fit of the combinatorial background.
Additionally, for the PHOS analysis, the signal region was excluded from the background fit and the signal was obtained by Gaussian integral instead of bin-by-bin counting. 
The material budget uncertainty accounts for a possible mismatch between the amount of material present in the ALICE detector and its implementation in \geant~3.
The material budget uncertainty for a conversion photon was studied in Ref.~\cite{Abelev:2012cn}, and found to be 4.5\% per photon. 
For the measurements involving the EMCal or the PHOS uncertainties of $3$ and $3.5$\%, respectively, were assigned for the material budget, which is dominated by the material of outer detectors positioned in front of calorimeter modules during data taking in 2010, as outlined in Ref.~\cite{Acharya:2017hyu}. 
The material uncertainty of the inner detectors is negligible for calorimeter photons due to the low conversion probability. 

The conversion and calorimeter photon reconstruction uncertainties were evaluated by independently varying the respective selection criteria given in \Sect{sec:photon}.
The conversion photon reconstruction uncertainty was found to be dominated by the reduced $\chi^2$ selection of the electron tracks and the requirement on the number of space points in the TPC for each track.
For EMCal related measurements, the cluster energy selection and the cluster shape have most influence on the uncertainty.
For PHOS, the photon reconstruction uncertainty was evaluated by variation of the track matching condition and cluster shape selection.
Uncertainties arising from the non-linearity and cluster energy scale of the respective calorimeters was taken into account by varying the scheme used to obtain the energy scale calibration and are included in the overall calorimeter photon reconstruction uncertainty.
Like the photon reconstruction uncertainties, the systematic uncertainties arising from the charged pion reconstruction were estimated by independent variation of the selection criteria given in \Sect{sec:eventtrack}.
To study the influence of in-bunch pileup on the measurement, the rejection criterium was loosened, resulting in a 0.5\% systematic uncertainty.
The systematic uncertainty due to the selection of neutral pion candidates was estimated by varying the invariant mass selection window.
For the PHOS measurement, the selection was additionally varied according to the $\pi^0$ candidates transverse momentum.
A detailed description of these sources of uncertainty is provided in \Refs{Acharya:2017hyu} and~\cite{omegaphosPN}.

\Table{tab:table_system_Pi0} also shows, for each method, the statistical uncertainty together with the total systematic uncertainty, which is obtained by adding the individual sources in quadrature.
In addition, the statistical and systematic uncertainties of the combined measurement are given, which were obtained taken into account correlations across the measurements as elaborated in the following section.

%%%%%%%%%%%%%%%%%%%%%%%%%%%%%%%%%%%%%%%%%%%%%%%%%%
\section{Results}
\label{sec:results}
%%%%%%%%%%%%%%%%%%%%%%%%%%%%%%%%%%%%%%%%%%%%%%%%%%
The fully corrected invariant cross sections of $\omega$ production were obtained for each reconstruction technique using
\begin{equation}
E\dv[3]{\sigma^{pp\rightarrow\omega + X}}{p} = \frac{1}{2\pi} \frac{1}{\text{\pT}}\cdot \frac{1}{\mathcal{L}_{\text{int}}} \cdot \frac{1}{A\cdot\varepsilon_{\text{rec.}}}\cdot\frac{1}{\text{BR}_{\omega\rightarrow\pi^+\pi^-\pi^0}}\cdot\frac{N^{\omega}}{\Delta y\Delta \text{\pT}}.
\end{equation}
Here, $\mathcal{L}_{\text{Int}}$ is the integrated luminosity given in \Sect{sec:eventtrack}, $\varepsilon_{\text{rec.}}$ and $A$ are the reconstruction efficiency and acceptance of the corresponding method and $\text{BR}=(89.3\pm0.6)\%$ is the branching ratio of the $\omega\rightarrow\pi^+\pi^-\pi^0$ decay~\cite{Tanabashi:2018oca}.
Moreover,  $N^{\omega}$ denotes the number of reconstructed $\omega$  mesons in the transverse momentum range $\Delta$\pT~and the given rapidity range $\Delta y$. 

The production cross sections were measured individually for each reconstruction method and then combined using \pT-dependent weights that are calculated according to the Best Linear Unbiased Estimate~(BLUE) algorithm~\cite{Valassi:2013bga}, which uses concepts that are routinely applied in statistical fields.
The combination took into account statistical and systematic uncertainties.
For the systematic uncertainties, the individual measurements are found to be correlated by about  30\%,  dominantly originating from the charged-pion selection and the material budget uncertainties.
These correlations were taken into account in the combination procedure.
The statistical and systematic uncertainties of the combined measurement are given in \Tab{tab:table_system_Pi0}.

\begin{figure}[th!]
	\centering
	\includegraphics[width=0.6\linewidth]{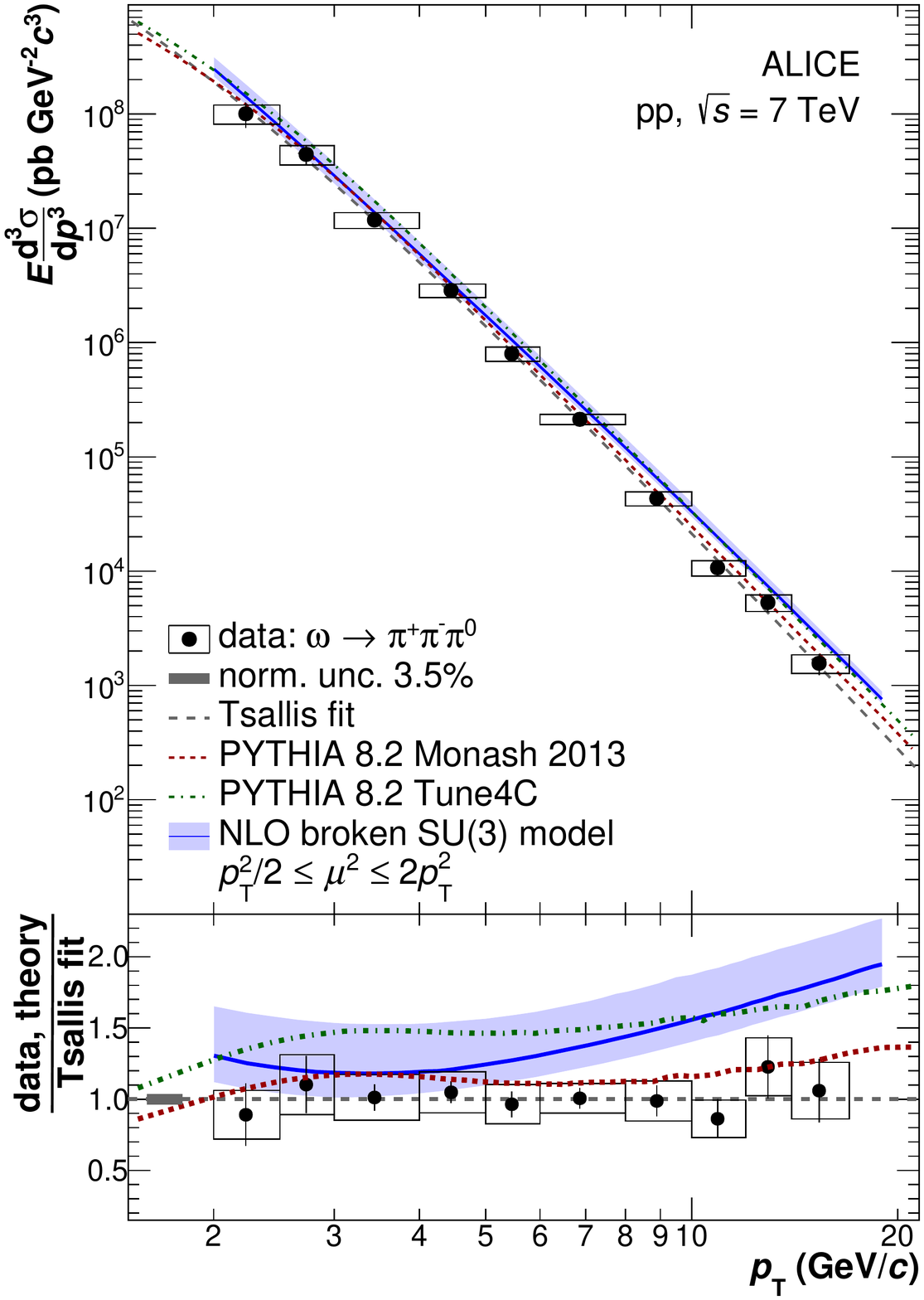}
	\caption{Invariant cross section of $\omega$ meson production in pp collisions at $\sqrt{s}=\SI{7}{TeV}$ compared to theoretical predictions.
		The statistical and systematic uncertainties are represented by vertical bars and boxes, respectively.
		A Levy-Tsallis function was used to parametrise the spectrum, where the obtained fit parameters are given in \Tab{tab:FitParam}.
		The predictions are obtained using \Pythia~8.2~\cite{Sjostrand:2014zea} with the Monash 2013~\cite{Skands:2014pea} and 4C~\cite{Tune4C} tunes.
		Furthermore, a NLO calculation~\cite{Saveetha:2017xmc} incorporating a model dedicated to describe vector-meson fragmentation is shown, where the band denotes the uncertainty of the scale $\mu$, which was used for factorisation, renormalisation and fragmentation.
		In the bottom panel, the ratios of the theoretical estimates to the Levy-Tsallis fit of the measurement are shown; the ratio of the data to the Levy-Tsallis fit is also presented.
	}
	\label{crosssection}
\end{figure}

The cross section of $\omega$ meson production for $2 < $\pT $<\SI{17}{GeV}/c$ at midrapidity in pp collisions at $\sqrt{s}=\SI{7}{TeV}$ is shown in \Fig{crosssection}.
It was fitted using a Levy-Tsallis function~\cite{Tsallis:1987eu} given by
\begin{equation}
E\frac{\mbox{d}^3\sigma}{\mbox{d}p^3} =\frac{C}{2\pi} \frac{(n-1)(n-2)}{nT [nT + m(n-2)]} \left(1 + \frac{m_{T}-m}{nT}\right)^{-n},
\label{eq:Tsallis}
\end{equation}
which describes the cross section over the whole measured transverse momentum range, as demonstrated in the lower panel of the figure.
The parameters $m$ and $m_{\text{\tiny T}} = \sqrt{m^2 + p_{\text{\tiny T}}^2}$ correspond to the particle mass and the transverse mass, respectively, while $C$, $T$ and $n$ are the free parameters of the Levy-Tsallis function.
\renewcommand{\arraystretch}{1.3}
\begin{table}[h!]
	\begin{center}
		\caption{Parameters and $\chi^2$/NDF of the fit to the $\omega$ invariant cross section using the Levy-Tsallis function~\cite{Tsallis:1987eu} from \Eq{eq:Tsallis}.}
		\label{tab:FitParam}
		\scalebox{0.92}{
			\begin{tabular}{c||ccccccc}
				\hline
				Levy-Tsallis & \multicolumn{2}{c}{$C$ ($\times10^{10}$ pb)} & $T$ (GeV) & \multicolumn{2}{c}{$n$} & $\chi^2$/NDF & NDF \\
				\hline\hline
				$\omega$ & \multicolumn{2}{c}{$4.01^{\pm2.47~\text{(stat)}}_{\pm3.41~\text{(tot)}}$}& $0.182^{\pm0.042~\text{(stat)}}_{ \pm0.061~\text{(tot)}}$ & \multicolumn{2}{c}{$6.46^{\pm0.37~\text{(stat)}}_{\pm0.55~\text{(tot)}}$} & $~^{0.45~\text{(stat)}}_{0.22~\text{(tot)}}$ & 7 \\
				\hline
		\end{tabular}}
	\end{center}
\end{table}
\renewcommand{\arraystretch}{1.0}

The values of the fit parameters and the reduced $\chi^2$ of the fit are given in \Tab{tab:FitParam}, where the fit was obtained using only statistical uncertainties, and using the systematic and statistical uncertainties of the measurement added in quadrature. 
To account for finite \pT-interval width, the combined cross section points were assigned to \pT~values shifted from the bin centre of the \pT~intervals according to the underlying spectrum~\cite{Lafferty:1994cj} described by a Levy-Tsallis function.
This correction resulted in a shift below 2\% in each \pT\ interval.

\begin{figure}[th!]
	\centering
	\includegraphics[width=0.5\linewidth]{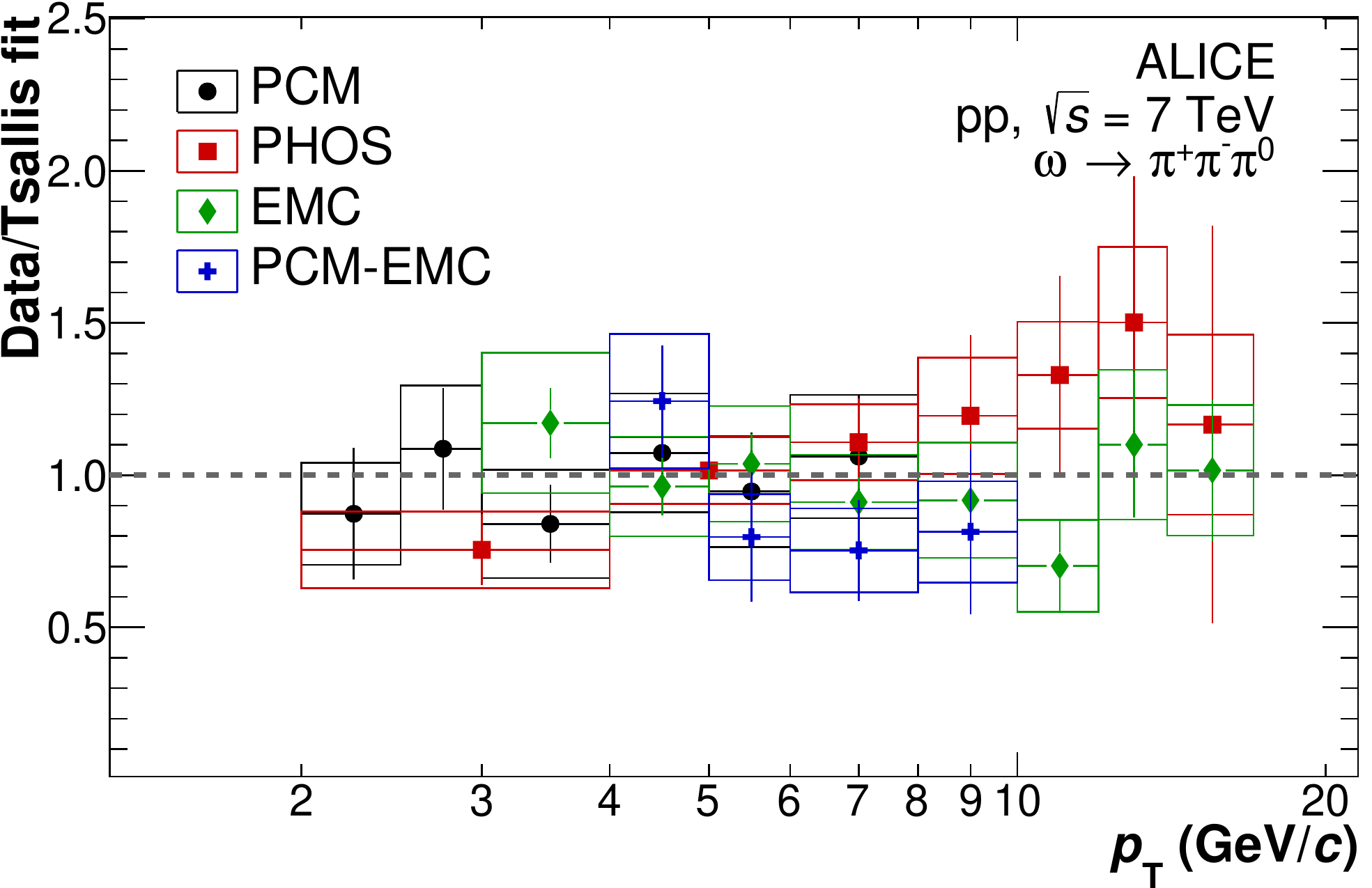}
	\caption{Ratios of the fully corrected $\omega$ spectra obtained with the individual reconstruction methods to the Levy-Tsallis fit of the combined spectrum, where the fit parameters are shown in \Tab{tab:FitParam}.
          The statistical and systematic uncertainties are represented by vertical bars and boxes, respectively.}
	\label{fig:omegaratioofindividualmeastocombfitpp7tev}
\end{figure}

\Figure{fig:omegaratioofindividualmeastocombfitpp7tev}, which shows the ratios of the cross sections for the individual reconstruction methods to the Levy-Tsallis fit of the combined measurement, demonstrates the agreement between all methods within the statistical and systematic uncertainties, justifying the combination of the individual results as discussed earlier.

The measured differential cross section of $\omega$ production is compared to several calculations in \Fig{crosssection}.
The ratio of each prediction to the Levy-Tsallis fit of the measurement is shown in the bottom panel of the figure. 
Two \Pythia~8.2~\cite{Sjostrand:2014zea} Monte Carlo event generator calculations were considered for comparison, which are based on the Monash 2013~\cite{Skands:2014pea} and the 4C~\cite{Tune4C} tunes, respectively.
The Monash 2013 tune describes the measurement over the full reported \pT~range within the uncertainties, while the Tune~4C overestimates the data by about 50\%.
The Monash 2013 tune includes more recent experimental results than Tune~4C and thus a more refined set of parameters.
In particular, the rate of light flavor vector meson production used in hadronisation process was revised and lowered, improving the description of $\omega$ meson yields~\cite{Skands:2014pea}.

The measurement is also compared to a next-to-leading order~(NLO) calculation using a model with broken SU(3) symmetry to describe vector meson production~\cite{Saveetha:2017xmc}, where the model parameters have been constrained using $\omega$ production data measured by PHENIX in pp collisions at $\sqrt{s}=\SI{200}{GeV}$~\cite{Adare:2010fe}.
The same scale $\mu=\text{\pT}$ was used for factorisation, renormalisation and fragmentation for the calculation and the shaded band reported in Fig.~\ref{crosssection} denotes the scale variation of $\text{\pT}^2/2\leq\mu^2\leq 2\text{\pT}^2$.
The calculation describes the measurement within the uncertainties below $\SI{6}{GeV}/c$, and overestimates the data by up to 50\% for higher \pT.

\begin{figure}[t!]
	\centering
	\includegraphics[width=0.7\linewidth]{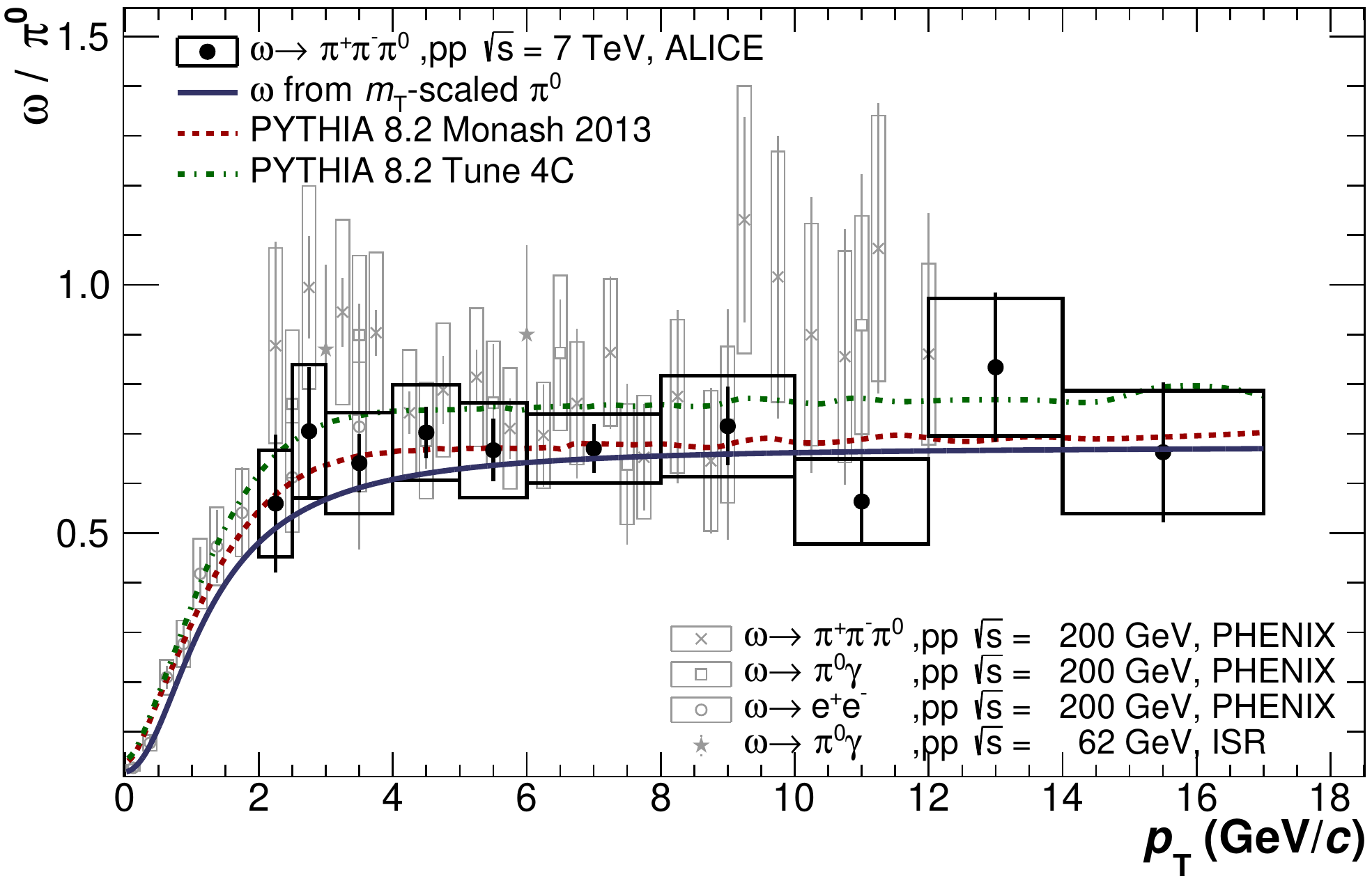}
	\caption{Ratio of $\omega/\pi^0$ production as a function of transverse momentum \pT~for pp collisions at $\sqrt{s}=\SI{7}{TeV}$ (black) compared to various lower collision energies ranging from $\sqrt{s}=62-\SI{200}{GeV}$~\cite{ISRomega,RHICOmega1,Adare:2010fe,RHICOmega2} (gray).
          In addition, \Pythia~predictions at $\sqrt{s}=\SI{7}{TeV}$ and the $\omega/\pi^0$ ratio obtained from \mt-scaling are shown with lines.}
	\label{omegapi0ratio}
\end{figure}

The ratio of $\omega$ relative to $\pi^0$ meson production is shown as a function of \pt\ in \Fig{omegapi0ratio}, where data points for the $\pi^0$ measurement were taken from Ref.~\cite{Abelev:2012cn}.
The ratio is observed to be constant above $\SI{2.5}{GeV}/c$ with a value of $C^{\omega/\pi^{0}}= 0.69  \pm 0.03 \text{~(stat)~} \pm 0.04 \text{~(sys)}$.
Within the uncertainties, the $\omega/\pi^0$ ratio is described by the \Pythia~predictions. 
Even though the Tune~4C  overestimates the $\omega$ production, it describes the $\omega/\pi^0$ ratio due to a similar overestimation of $\pi^0$ production, which was reported in Ref.~\cite{Acharya:2017tlv}.

The measured $\omega/\pi^0$ ratio at $\sqrt{s}=\SI{7}{TeV}$ is compared to data from lower collision energies at $\sqrt{s} = 62$~\cite{ISRomega} and \SI{200}{GeV}~\cite{Adare:2010fe,RHICOmega1,RHICOmega2}.
The  $\omega/\pi^0$  ratios measured at the different collision energies agree within the uncertainties.
In order to test the validity of \mT-scaling, the Levy-Tsallis parametrisation $f_{\pi^0}(p_{\text{\tiny T,}\pi^0})$ of the $\pi^0$  spectrum reported in Ref.~\cite{Abelev:2012cn} was scaled using the ratio $C^{\omega/\pi^{0}}= 0.67$, following the procedure discussed in detail in Ref.~\cite{Altenkamper:2017qot}.
The scaled parametrisation $f_\omega(p_{\text{\tiny T,}\omega})$ was used to calculate the $\omega/\pi^0$ ratio via $f_\omega(p_{\text{\tiny T,}\omega})/f_{\pi^0}(p_{\text{\tiny T,}\omega})$, where the relation $p_{\text{\tiny T,}\omega}^2 + m_{0,\omega}^2 =  p_{\text{\tiny T,}\pi^0}^2 + m_{0,\pi^0}^2$ was used to ensure the evaluation of both spectra at the same transverse mass.
The obtained \mT-scaling prediction of the $\omega/\pi^0$ ratio is shown in \Fig{omegapi0ratio} and found to be consistent with the measurement. 
Unlike in the case of the $\eta/\pi^0$ ratio  measured at $\sqrt{s} = 2.76$, $7$ and \SI{8}{TeV}~\cite{Abelev:2012cn,Acharya:2017hyu,Acharya:2017tlv}, where a violation of \mt-scaling was observed below \SI{3.5}{GeV}, no such violation is observed within the uncertainties for the $\omega$ meson in the entire measured momentum range. However, while the measurement is compatible with the \mT-scaling prediction at low-\pT, the sensitivity of the measurement to a possible \mt-scaling violation is limited by the uncertainties and \pT~reach. Here, future studies with increased precision could provide further insights and more stringent tests of \mT-scaling for low-\pT~$\omega$ mesons.
Interestingly, the \Pythia~calculations and the \mT-scaled prediction both describe the $\omega/\pi^0$ ratio at lower collision energies even below $\pT=\SI{2}{GeV}/c$, suggesting a universal feature of meson production.

%%%%%%%%%%%%%%%%%%%%%%%%%%%%%%%%%%%%%%%%%%%%%%%%%%
\section{Conclusion}
\label{sec:conclusion}
%%%%%%%%%%%%%%%%%%%%%%%%%%%%%%%%%%%%%%%%%%%%%%%%%%
The invariant differential cross section of $\omega$ meson production at midrapidity in pp collisions $\sqrt{s}=\SI{7}{TeV}$ was measured with the ALICE detector, covering a transverse-momentum range of 2 to $\SI{17}{GeV}/c$.
Within the uncertainties, \Pythia~8.2 predictions for the Monash 2013 tune describes the measurement over the whole \pT~range, while Tune~4C overestimates the data by about 50\%.
A NLO calculation using a model dedicated to describing fragmentation into the entire vector meson nonet describes the data below $\SI{6}{GeV}/c$, while it overestimates the data by up to 50\% at higher $\pt$. 
Above  $\SI{2.5}{GeV}/c$ the $\omega/\pi^0$ ratio is found to be constant with a value of $C^{\omega/\pi^{0}}= 0.67  \pm 0.03 \text{~(stat)~} \pm 0.04 \text{~(sys)}$ and agrees with measurements at lower collision energies and with \Pythia~predictions over the whole reported \pT~range.
Within the uncertainties, the \mT-scaling prediction for $\omega$ mesons is consistent with the measured spectrum above $\SI{2}{GeV}/c$.

%%%%%%%%%%%%%%%%%%%%%%%%%%%%%%%%%%%%%%%%%%%%%%%%%
\newenvironment{acknowledgement}{\relax}{\relax}
%%%%%%%%%%%%%%%%%%%%%%%%%%%%%%%%%%%%%%%%%%%%%%%%%%
\begin{acknowledgement}
\section*{Acknowledgments}
We thank D.~Indumathi for providing the NLO calculations.

% Version: 2020-06-16

The ALICE Collaboration would like to thank all its engineers and technicians for their invaluable contributions to the construction of the experiment and the CERN accelerator teams for the outstanding performance of the LHC complex.
The ALICE Collaboration gratefully acknowledges the resources and support provided by all Grid centres and the Worldwide LHC Computing Grid (WLCG) collaboration.
The ALICE Collaboration acknowledges the following funding agencies for their support in building and running the ALICE detector:
A. I. Alikhanyan National Science Laboratory (Yerevan Physics Institute) Foundation (ANSL), State Committee of Science and World Federation of Scientists (WFS), Armenia;
Austrian Academy of Sciences, Austrian Science Fund (FWF): [M 2467-N36] and Nationalstiftung f\"{u}r Forschung, Technologie und Entwicklung, Austria;
Ministry of Communications and High Technologies, National Nuclear Research Center, Azerbaijan;
Conselho Nacional de Desenvolvimento Cient\'{\i}fico e Tecnol\'{o}gico (CNPq), Financiadora de Estudos e Projetos (Finep), Funda\c{c}\~{a}o de Amparo \`{a} Pesquisa do Estado de S\~{a}o Paulo (FAPESP) and Universidade Federal do Rio Grande do Sul (UFRGS), Brazil;
Ministry of Education of China (MOEC) , Ministry of Science \& Technology of China (MSTC) and National Natural Science Foundation of China (NSFC), China;
Ministry of Science and Education and Croatian Science Foundation, Croatia;
Centro de Aplicaciones Tecnol\'{o}gicas y Desarrollo Nuclear (CEADEN), Cubaenerg\'{\i}a, Cuba;
Ministry of Education, Youth and Sports of the Czech Republic, Czech Republic;
The Danish Council for Independent Research | Natural Sciences, the VILLUM FONDEN and Danish National Research Foundation (DNRF), Denmark;
Helsinki Institute of Physics (HIP), Finland;
Commissariat \`{a} l'Energie Atomique (CEA) and Institut National de Physique Nucl\'{e}aire et de Physique des Particules (IN2P3) and Centre National de la Recherche Scientifique (CNRS), France;
Bundesministerium f\"{u}r Bildung und Forschung (BMBF) and GSI Helmholtzzentrum f\"{u}r Schwerionenforschung GmbH, Germany;
General Secretariat for Research and Technology, Ministry of Education, Research and Religions, Greece;
National Research, Development and Innovation Office, Hungary;
Department of Atomic Energy Government of India (DAE), Department of Science and Technology, Government of India (DST), University Grants Commission, Government of India (UGC) and Council of Scientific and Industrial Research (CSIR), India;
Indonesian Institute of Science, Indonesia;
Centro Fermi - Museo Storico della Fisica e Centro Studi e Ricerche Enrico Fermi and Istituto Nazionale di Fisica Nucleare (INFN), Italy;
Institute for Innovative Science and Technology , Nagasaki Institute of Applied Science (IIST), Japanese Ministry of Education, Culture, Sports, Science and Technology (MEXT) and Japan Society for the Promotion of Science (JSPS) KAKENHI, Japan;
Consejo Nacional de Ciencia (CONACYT) y Tecnolog\'{i}a, through Fondo de Cooperaci\'{o}n Internacional en Ciencia y Tecnolog\'{i}a (FONCICYT) and Direcci\'{o}n General de Asuntos del Personal Academico (DGAPA), Mexico;
Nederlandse Organisatie voor Wetenschappelijk Onderzoek (NWO), Netherlands;
The Research Council of Norway, Norway;
Commission on Science and Technology for Sustainable Development in the South (COMSATS), Pakistan;
Pontificia Universidad Cat\'{o}lica del Per\'{u}, Peru;
Ministry of Science and Higher Education, National Science Centre and WUT ID-UB, Poland;
Korea Institute of Science and Technology Information and National Research Foundation of Korea (NRF), Republic of Korea;
Ministry of Education and Scientific Research, Institute of Atomic Physics and Ministry of Research and Innovation and Institute of Atomic Physics, Romania;
Joint Institute for Nuclear Research (JINR), Ministry of Education and Science of the Russian Federation, National Research Centre Kurchatov Institute, Russian Science Foundation and Russian Foundation for Basic Research, Russia;
Ministry of Education, Science, Research and Sport of the Slovak Republic, Slovakia;
National Research Foundation of South Africa, South Africa;
Swedish Research Council (VR) and Knut \& Alice Wallenberg Foundation (KAW), Sweden;
European Organization for Nuclear Research, Switzerland;
Suranaree University of Technology (SUT), National Science and Technology Development Agency (NSDTA) and Office of the Higher Education Commission under NRU project of Thailand, Thailand;
Turkish Atomic Energy Agency (TAEK), Turkey;
National Academy of  Sciences of Ukraine, Ukraine;
Science and Technology Facilities Council (STFC), United Kingdom;
National Science Foundation of the United States of America (NSF) and United States Department of Energy, Office of Nuclear Physics (DOE NP), United States of America.    %%%%%%% done by webmaster team
\end{acknowledgement}

%%%%%%%%%%%%%%%%%%%%%%%%%%%%%%%%%%%%%%%%%%%%%%%%%%
\bibliographystyle{utphys}
\bibliography{biblio}{}
%%%%%%%%%%%%%%%%%%%%%%%%%%%%%%%%%%%%%%%%%%%%%%%%%%
\newpage
\appendix
%%%%%%%%%%%%%%%%%%%%%%%%%%%%%%%%%%%%%%%%%%%%%%%%%%
\section{The ALICE Collaboration}
\label{app:collab}
% Collaboration: CERN-LHC-ALICE
% Generation Date is 2020-06-16

% How to use:
%%%%%%%%% appendix with author list
%\appendix
%\section{The ALICE Collaboration}
%\label{app:collab}
%\input{Alice_Authorslist_XXXX-Axx-XX.tex}
\begingroup
\small
\begin{flushleft}
S.~Acharya\Irefn{org141}\And 
D.~Adamov\'{a}\Irefn{org95}\And 
A.~Adler\Irefn{org74}\And 
J.~Adolfsson\Irefn{org81}\And 
M.M.~Aggarwal\Irefn{org100}\And 
S.~Agha\Irefn{org14}\And 
G.~Aglieri Rinella\Irefn{org34}\And 
M.~Agnello\Irefn{org30}\And 
N.~Agrawal\Irefn{org10}\textsuperscript{,}\Irefn{org54}\And 
Z.~Ahammed\Irefn{org141}
% ...
\And
\renewcommand\labelenumi{\textsuperscript{\theenumi}~}

\section*{Affiliation notes}
\renewcommand\theenumi{\roman{enumi}}
\begin{Authlist}
\item \Adef{org*}Deceased
\item \Adef{orgI}Italian National Agency for New Technologies, Energy and Sustainable Economic Development (ENEA), Bologna, Italy
\item \Adef{orgII}Dipartimento DET del Politecnico di Torino, Turin, Italy
\item \Adef{orgIII}M.V. Lomonosov Moscow State University, D.V. Skobeltsyn Institute of Nuclear, Physics, Moscow, Russia
\item \Adef{orgIV}Department of Applied Physics, Aligarh Muslim University, Aligarh, India
\item \Adef{orgV}Institute of Theoretical Physics, University of Wroclaw, Poland
\end{Authlist}

\section*{Collaboration Institutes}
\renewcommand\theenumi{\arabic{enumi}~}
\begin{Authlist}
\item \Idef{org1}A.I. Alikhanyan National Science Laboratory (Yerevan Physics Institute) Foundation, Yerevan, Armenia
\item \Idef{org2}Bogolyubov Institute for Theoretical Physics, National Academy of Sciences of Ukraine, Kiev, Ukraine
\item \Idef{org3}Bose Institute, Department of Physics  and Centre for Astroparticle Physics and Space Science (CAPSS), Kolkata, India
\item \Idef{org4}Budker Institute for Nuclear Physics, Novosibirsk, Russia
\item \Idef{org5}California Polytechnic State University, San Luis Obispo, California, United States
\item \Idef{org6}Central China Normal University, Wuhan, China
\item \Idef{org7}Centre de Calcul de l'IN2P3, Villeurbanne, Lyon, France
\item \Idef{org8}Centro de Aplicaciones Tecnol\'{o}gicas y Desarrollo Nuclear (CEADEN), Havana, Cuba
\item \Idef{org9}Centro de Investigaci\'{o}n y de Estudios Avanzados (CINVESTAV), Mexico City and M\'{e}rida, Mexico
\item \Idef{org10}Centro Fermi - Museo Storico della Fisica e Centro Studi e Ricerche ``Enrico Fermi', Rome, Italy
\item \Idef{org11}Chicago State University, Chicago, Illinois, United States
\item \Idef{org12}China Institute of Atomic Energy, Beijing, China
\item \Idef{org13}Comenius University Bratislava, Faculty of Mathematics, Physics and Informatics, Bratislava, Slovakia
\item \Idef{org14}COMSATS University Islamabad, Islamabad, Pakistan
\item \Idef{org15}Creighton University, Omaha, Nebraska, United States
\item \Idef{org16}Department of Physics, Aligarh Muslim University, Aligarh, India
\item \Idef{org17}Department of Physics, Pusan National University, Pusan, Republic of Korea
\item \Idef{org18}Department of Physics, Sejong University, Seoul, Republic of Korea
\item \Idef{org19}Department of Physics, University of California, Berkeley, California, United States
\item \Idef{org20}Department of Physics, University of Oslo, Oslo, Norway
\item \Idef{org21}Department of Physics and Technology, University of Bergen, Bergen, Norway
\item \Idef{org22}Dipartimento di Fisica dell'Universit\`{a} 'La Sapienza' and Sezione INFN, Rome, Italy
\item \Idef{org23}Dipartimento di Fisica dell'Universit\`{a} and Sezione INFN, Cagliari, Italy
\item \Idef{org24}Dipartimento di Fisica dell'Universit\`{a} and Sezione INFN, Trieste, Italy
\item \Idef{org25}Dipartimento di Fisica dell'Universit\`{a} and Sezione INFN, Turin, Italy
\item \Idef{org26}Dipartimento di Fisica e Astronomia dell'Universit\`{a} and Sezione INFN, Bologna, Italy
\item \Idef{org27}Dipartimento di Fisica e Astronomia dell'Universit\`{a} and Sezione INFN, Catania, Italy
\item \Idef{org28}Dipartimento di Fisica e Astronomia dell'Universit\`{a} and Sezione INFN, Padova, Italy
\item \Idef{org29}Dipartimento di Fisica `E.R.~Caianiello' dell'Universit\`{a} and Gruppo Collegato INFN, Salerno, Italy
\item \Idef{org30}Dipartimento DISAT del Politecnico and Sezione INFN, Turin, Italy
\item \Idef{org31}Dipartimento di Scienze e Innovazione Tecnologica dell'Universit\`{a} del Piemonte Orientale and INFN Sezione di Torino, Alessandria, Italy
\item \Idef{org32}Dipartimento di Scienze MIFT, Universit\`{a} di Messina, Messina, Italy
\item \Idef{org33}Dipartimento Interateneo di Fisica `M.~Merlin' and Sezione INFN, Bari, Italy
\item \Idef{org34}European Organization for Nuclear Research (CERN), Geneva, Switzerland
\item \Idef{org35}Faculty of Electrical Engineering, Mechanical Engineering and Naval Architecture, University of Split, Split, Croatia
\item \Idef{org36}Faculty of Engineering and Science, Western Norway University of Applied Sciences, Bergen, Norway
\item \Idef{org37}Faculty of Nuclear Sciences and Physical Engineering, Czech Technical University in Prague, Prague, Czech Republic
\item \Idef{org38}Faculty of Science, P.J.~\v{S}af\'{a}rik University, Ko\v{s}ice, Slovakia
\item \Idef{org39}Frankfurt Institute for Advanced Studies, Johann Wolfgang Goethe-Universit\"{a}t Frankfurt, Frankfurt, Germany
\item \Idef{org40}Fudan University, Shanghai, China
\item \Idef{org41}Gangneung-Wonju National University, Gangneung, Republic of Korea
\item \Idef{org42}Gauhati University, Department of Physics, Guwahati, India
\item \Idef{org43}Helmholtz-Institut f\"{u}r Strahlen- und Kernphysik, Rheinische Friedrich-Wilhelms-Universit\"{a}t Bonn, Bonn, Germany
\item \Idef{org44}Helsinki Institute of Physics (HIP), Helsinki, Finland
\item \Idef{org45}High Energy Physics Group,  Universidad Aut\'{o}noma de Puebla, Puebla, Mexico
\item \Idef{org46}Hiroshima University, Hiroshima, Japan
\item \Idef{org47}Hochschule Worms, Zentrum  f\"{u}r Technologietransfer und Telekommunikation (ZTT), Worms, Germany
\item \Idef{org48}Horia Hulubei National Institute of Physics and Nuclear Engineering, Bucharest, Romania
\item \Idef{org49}Indian Institute of Technology Bombay (IIT), Mumbai, India
\item \Idef{org50}Indian Institute of Technology Indore, Indore, India
\item \Idef{org51}Indonesian Institute of Sciences, Jakarta, Indonesia
\item \Idef{org52}INFN, Laboratori Nazionali di Frascati, Frascati, Italy
\item \Idef{org53}INFN, Sezione di Bari, Bari, Italy
\item \Idef{org54}INFN, Sezione di Bologna, Bologna, Italy
\item \Idef{org55}INFN, Sezione di Cagliari, Cagliari, Italy
\item \Idef{org56}INFN, Sezione di Catania, Catania, Italy
\item \Idef{org57}INFN, Sezione di Padova, Padova, Italy
\item \Idef{org58}INFN, Sezione di Roma, Rome, Italy
\item \Idef{org59}INFN, Sezione di Torino, Turin, Italy
\item \Idef{org60}INFN, Sezione di Trieste, Trieste, Italy
\item \Idef{org61}Inha University, Incheon, Republic of Korea
\item \Idef{org62}Institute for Nuclear Research, Academy of Sciences, Moscow, Russia
\item \Idef{org63}Institute for Subatomic Physics, Utrecht University/Nikhef, Utrecht, Netherlands
\item \Idef{org64}Institute of Experimental Physics, Slovak Academy of Sciences, Ko\v{s}ice, Slovakia
\item \Idef{org65}Institute of Physics, Homi Bhabha National Institute, Bhubaneswar, India
\item \Idef{org66}Institute of Physics of the Czech Academy of Sciences, Prague, Czech Republic
\item \Idef{org67}Institute of Space Science (ISS), Bucharest, Romania
\item \Idef{org68}Institut f\"{u}r Kernphysik, Johann Wolfgang Goethe-Universit\"{a}t Frankfurt, Frankfurt, Germany
\item \Idef{org69}Instituto de Ciencias Nucleares, Universidad Nacional Aut\'{o}noma de M\'{e}xico, Mexico City, Mexico
\item \Idef{org70}Instituto de F\'{i}sica, Universidade Federal do Rio Grande do Sul (UFRGS), Porto Alegre, Brazil
\item \Idef{org71}Instituto de F\'{\i}sica, Universidad Nacional Aut\'{o}noma de M\'{e}xico, Mexico City, Mexico
\item \Idef{org72}iThemba LABS, National Research Foundation, Somerset West, South Africa
\item \Idef{org73}Jeonbuk National University, Jeonju, Republic of Korea
\item \Idef{org74}Johann-Wolfgang-Goethe Universit\"{a}t Frankfurt Institut f\"{u}r Informatik, Fachbereich Informatik und Mathematik, Frankfurt, Germany
\item \Idef{org75}Joint Institute for Nuclear Research (JINR), Dubna, Russia
\item \Idef{org76}Korea Institute of Science and Technology Information, Daejeon, Republic of Korea
\item \Idef{org77}KTO Karatay University, Konya, Turkey
\item \Idef{org78}Laboratoire de Physique des 2 Infinis, Ir\`{e}ne Joliot-Curie, Orsay, France
\item \Idef{org79}Laboratoire de Physique Subatomique et de Cosmologie, Universit\'{e} Grenoble-Alpes, CNRS-IN2P3, Grenoble, France
\item \Idef{org80}Lawrence Berkeley National Laboratory, Berkeley, California, United States
\item \Idef{org81}Lund University Department of Physics, Division of Particle Physics, Lund, Sweden
\item \Idef{org82}Nagasaki Institute of Applied Science, Nagasaki, Japan
\item \Idef{org83}Nara Women{'}s University (NWU), Nara, Japan
\item \Idef{org84}National and Kapodistrian University of Athens, School of Science, Department of Physics , Athens, Greece
\item \Idef{org85}National Centre for Nuclear Research, Warsaw, Poland
\item \Idef{org86}National Institute of Science Education and Research, Homi Bhabha National Institute, Jatni, India
\item \Idef{org87}National Nuclear Research Center, Baku, Azerbaijan
\item \Idef{org88}National Research Centre Kurchatov Institute, Moscow, Russia
\item \Idef{org89}Niels Bohr Institute, University of Copenhagen, Copenhagen, Denmark
\item \Idef{org90}Nikhef, National institute for subatomic physics, Amsterdam, Netherlands
\item \Idef{org91}NRC Kurchatov Institute IHEP, Protvino, Russia
\item \Idef{org92}NRC \guillemotleft Kurchatov\guillemotright  Institute - ITEP, Moscow, Russia
\item \Idef{org93}NRNU Moscow Engineering Physics Institute, Moscow, Russia
\item \Idef{org94}Nuclear Physics Group, STFC Daresbury Laboratory, Daresbury, United Kingdom
\item \Idef{org95}Nuclear Physics Institute of the Czech Academy of Sciences, \v{R}e\v{z} u Prahy, Czech Republic
\item \Idef{org96}Oak Ridge National Laboratory, Oak Ridge, Tennessee, United States
\item \Idef{org97}Ohio State University, Columbus, Ohio, United States
\item \Idef{org98}Petersburg Nuclear Physics Institute, Gatchina, Russia
\item \Idef{org99}Physics department, Faculty of science, University of Zagreb, Zagreb, Croatia
\item \Idef{org100}Physics Department, Panjab University, Chandigarh, India
\item \Idef{org101}Physics Department, University of Jammu, Jammu, India
\item \Idef{org102}Physics Department, University of Rajasthan, Jaipur, India
\item \Idef{org103}Physikalisches Institut, Eberhard-Karls-Universit\"{a}t T\"{u}bingen, T\"{u}bingen, Germany
\item \Idef{org104}Physikalisches Institut, Ruprecht-Karls-Universit\"{a}t Heidelberg, Heidelberg, Germany
\item \Idef{org105}Physik Department, Technische Universit\"{a}t M\"{u}nchen, Munich, Germany
\item \Idef{org106}Politecnico di Bari, Bari, Italy
\item \Idef{org107}Research Division and ExtreMe Matter Institute EMMI, GSI Helmholtzzentrum f\"ur Schwerionenforschung GmbH, Darmstadt, Germany
\item \Idef{org108}Rudjer Bo\v{s}kovi\'{c} Institute, Zagreb, Croatia
\item \Idef{org109}Russian Federal Nuclear Center (VNIIEF), Sarov, Russia
\item \Idef{org110}Saha Institute of Nuclear Physics, Homi Bhabha National Institute, Kolkata, India
\item \Idef{org111}School of Physics and Astronomy, University of Birmingham, Birmingham, United Kingdom
\item \Idef{org112}Secci\'{o}n F\'{\i}sica, Departamento de Ciencias, Pontificia Universidad Cat\'{o}lica del Per\'{u}, Lima, Peru
\item \Idef{org113}St. Petersburg State University, St. Petersburg, Russia
\item \Idef{org114}Stefan Meyer Institut f\"{u}r Subatomare Physik (SMI), Vienna, Austria
\item \Idef{org115}SUBATECH, IMT Atlantique, Universit\'{e} de Nantes, CNRS-IN2P3, Nantes, France
\item \Idef{org116}Suranaree University of Technology, Nakhon Ratchasima, Thailand
\item \Idef{org117}Technical University of Ko\v{s}ice, Ko\v{s}ice, Slovakia
\item \Idef{org118}The Henryk Niewodniczanski Institute of Nuclear Physics, Polish Academy of Sciences, Cracow, Poland
\item \Idef{org119}The University of Texas at Austin, Austin, Texas, United States
\item \Idef{org120}Universidad Aut\'{o}noma de Sinaloa, Culiac\'{a}n, Mexico
\item \Idef{org121}Universidade de S\~{a}o Paulo (USP), S\~{a}o Paulo, Brazil
\item \Idef{org122}Universidade Estadual de Campinas (UNICAMP), Campinas, Brazil
\item \Idef{org123}Universidade Federal do ABC, Santo Andre, Brazil
\item \Idef{org124}University of Cape Town, Cape Town, South Africa
\item \Idef{org125}University of Houston, Houston, Texas, United States
\item \Idef{org126}University of Jyv\"{a}skyl\"{a}, Jyv\"{a}skyl\"{a}, Finland
\item \Idef{org127}University of Liverpool, Liverpool, United Kingdom
\item \Idef{org128}University of Science and Technology of China, Hefei, China
\item \Idef{org129}University of South-Eastern Norway, Tonsberg, Norway
\item \Idef{org130}University of Tennessee, Knoxville, Tennessee, United States
\item \Idef{org131}University of the Witwatersrand, Johannesburg, South Africa
\item \Idef{org132}University of Tokyo, Tokyo, Japan
\item \Idef{org133}University of Tsukuba, Tsukuba, Japan
\item \Idef{org134}Universit\'{e} Clermont Auvergne, CNRS/IN2P3, LPC, Clermont-Ferrand, France
\item \Idef{org135}Universit\'{e} de Lyon, Universit\'{e} Lyon 1, CNRS/IN2P3, IPN-Lyon, Villeurbanne, Lyon, France
\item \Idef{org136}Universit\'{e} de Strasbourg, CNRS, IPHC UMR 7178, F-67000 Strasbourg, France, Strasbourg, France
\item \Idef{org137}Universit\'{e} Paris-Saclay Centre d'Etudes de Saclay (CEA), IRFU, D\'{e}partment de Physique Nucl\'{e}aire (DPhN), Saclay, France
\item \Idef{org138}Universit\`{a} degli Studi di Foggia, Foggia, Italy
\item \Idef{org139}Universit\`{a} degli Studi di Pavia, Pavia, Italy
\item \Idef{org140}Universit\`{a} di Brescia, Brescia, Italy
\item \Idef{org141}Variable Energy Cyclotron Centre, Homi Bhabha National Institute, Kolkata, India
\item \Idef{org142}Warsaw University of Technology, Warsaw, Poland
\item \Idef{org143}Wayne State University, Detroit, Michigan, United States
\item \Idef{org144}Westf\"{a}lische Wilhelms-Universit\"{a}t M\"{u}nster, Institut f\"{u}r Kernphysik, M\"{u}nster, Germany
\item \Idef{org145}Wigner Research Centre for Physics, Budapest, Hungary
\item \Idef{org146}Yale University, New Haven, Connecticut, United States
\item \Idef{org147}Yonsei University, Seoul, Republic of Korea
\end{Authlist}
\endgroup
  %%%%%%% done by webmaster team
\end{document}